# Comparison of Chemical Evolution Models for the Galactic Disk

Monica Tosi

*Osservatorio Astronomico di Bologna, Bologna, Italy*

**Abstract.** The *best* chemical evolution models for the galactic disk computed by different groups with different assumptions are compared with each other and with the observational constraints. Differences and similarities between the models are discussed, emphasizing the importance of testing not only the solar neighbourhood but the whole disk. The history of the metal abundance gradients is found to be very sensitive to the different model assumptions on the star formation and infall rates, and can therefore provide important clues on the disk evolutionary scenario. All the examined models indicate that the deuterium depletion from its primordial abundance to the present one has been at most a factor of 3 and that the stellar nucleosynthesis of some elements (e.g. $^3$He, $^{13}$C and, possibly, $^{18}$O) requires some revisions.

## 1. Introduction

Modelling the chemical evolution of galaxies has proven to be a fundamental tool to study not only the evolution of these systems but also clusters of galaxies, intergalactic medium, Big Bang nucleosynthesis, etc. The results of these models, however, are not always taken in due consideration, with the argument that there are too many free parameters involved. The organizers of this meeting have thus suggested to compare chemical evolution models built by different groups to check what the different approaches lead to predict. To this aim, they have asked to several people working in the field to provide any possible information on their models and on the corresponding predictions for the time and space behaviours in the disk of our Galaxy of D, $^3$He, C, N and O isotopes, and Fe. The persons who have positively answered to such request (see Table 1) have models consistent with the majority of the observational constraints available for the galactic disk, and have sent the data relative to their *best* case, i.e the model with predictions in better agreement with the various data sets. This implies that the models compared here have been computed not only with different approaches and different numerical codes, but also with different assumptions on several parameters. What the examined models have in common is that they all take into account the finite stellar lifetimes, thus avoiding the instantaneous recycling approximation, and they all do not include dynamics.

The major assumptions made by each author for his/her best model are summarized in Table 1 (where all times are in Gyr and all distances in kpc). Most authors assume a disk age around 13 Gyr (second column) and a solar galactocentric distance around 8 kpc (third column). The latter value does not significantly affect the model predictions, but it should be borne in mind to scale everything properly in the comparison between predicted values and corresponding observational data. For a more direct comparison of the various results, hereinafter both model predictions and empirical quantities have all been scaled to a solar galactocentric distance of 8 kpc.

Table 1.  Input Parameters of the *Best* Models.

| Author[a]  | T    | $R_\odot$ | IMF[b] | SFR                                    | $\theta$     | yields[c]     |
|------------|------|-----------|--------|----------------------------------------|--------------|---------------|
| Carigi     | 13   | 8         | Kr93   | $A\Sigma_g^{1.4}(R)\Sigma_{tot}^{0.4}(R)$ | 0.6R-1.8     | RV81+M92      |
| Ferrini    | 13   | 8         | Fe90   | multiphase                             | 1-2          | RV81+WW86     |
| Matteucci  | 13   | 10        | Sc86   | $A\Sigma_g^{1.1}(R)\Sigma_{tot}^{0.1}(R)$ | 0.464R-1.59  | RV81+WW94     |
| Prantzos-a | 13.5 | 8.5       | Kr93   | $0.3\Sigma_g(R)/(R/R_\odot)$           | 3            | M95+RV81+WW93 |
| Timmes     | 15   | 8.5       | Salp   | $A\Sigma_g^2(R)$                       | 4            | RV81+WW94     |
| Tosi-1     | 13   | 8         | Ti80   | $A(R)\exp(-t/\tau)$  $\tau=15$         | $\infty$     | RV81+CC79     |

[a]References for these models are: Carigi 1994 and 1996; Pardi & Ferrini 1994, Ferrini et al 1994, and Galli et al 1995; Matteucci and François 1989 and Chiappini & Matteucci 1996; Prantzos & Aubert 1995, Prantzos 1996, and Prantzos et al 1996; Timmes et al 1995; Tosi 1988, Giovagnoli & Tosi 1995, and Dearborn et al 1996

[b]IMF references are: Kroupa et al 1993, Ferrini et al 1990, Scalo 1986, Salpeter 1955 and Tinsley 1980

[c]References for the adopted yields are: Marigo et al 1995 and Renzini & Voli 1981 for low and intermediate mass stars, Maeder 1992, Woosley & Weaver 1986, Weaver & Woosley 1993, Woosley & Weaver 1994, and Chiosi & Caimmi 1979 for massive stars. For Fe, Matteucci & Greggio's 1986 prescriptions are usually adopted.

There is apparently a large variety of initial mass functions (IMF) suitable for the galactic disk (fourth column). What should be emphasized is that, with only one exception, all authors find that an IMF with massive stars slope steeper than Salpeter's seems more appropriate for the disk stellar populations (recall that Salpeter derived his function only from low and intermediate mass stars).

Since the physical mechanism for stellar formation is still uncertain, we assume for the star formation rate (SFR) analytical functions (fifth column in Table 1) somehow inferred from empirical considerations. Most often the SFR is assumed to depend on the gas density $\Sigma_g(R)$, the only exceptions being Ferrini's and Tosi's models. In the latter case, the SFR is assumed to be simply an exponentially decreasing function of time (but with the coefficient A(R) derived from the present gas and total mass densities, $\Sigma_g(R,T)$ and $\Sigma_{tot}(R,T)$, respectively). In Ferrini's case, the star formation law is much more sophisticated, since he takes into account the different gas phases and derives the SFR from the probability of cloud collisions.

All the examined models assume that the disk has accreted metal poor gas from outside. The adopted infall e-folding times $\theta$ (listed in the sixth column of Table 1) range from a minimum of 1-2 Gyr to a maximum of infinity (i.e. constant infall). Such a large difference in the adopted infall time scales corresponds to different scenarios: small $\theta$ imply that the disk accretes gas only from the halo collapse, which lasts at most a few Gyr; large $\theta$ imply that the infalling gas comes not only from the halo but also from the intergalactic medium, for instance through the Magellanic Stream, which is an ongoing, observed phe-nomenon. Some authors adopt the same infall e-folding time for all disk regions, whereas others prefer an inside-out disk formation, with $\theta$ proportional to the galactocentric distance R.

The last column of Table 1 lists the adopted stellar yields. In the range of low and intermediate mass stars we all adopt Renzini and Voli's (1981) results, which are still the only complete set available for these masses. New yields, but only for stars below 4 $M_\odot$, have been computed by Marigo et al. (1996) and adopted by Prantzos. For massive stars, various groups adopt different versions of Woosley & Weaver's yields, Carigi adopts the recent values obtained by Maeder (1992) taking into account the strong effect of the initial metallicity on stellar yields, and Tosi uses the classic data by Arnett (1978) as revised by Chiosi and Caimmi (1979) to take stellar mass loss into account.

Table 2 shows some relevant quantities resulting from the models described in Table 1. The SFRs (in $M_\odot pc^{-2} Gyr^{-1}$) predicted at the solar distance at the time of sun formation (4.5 Gyr ago) and at the present time are given in columns 2 and 3 respectively; the corresponding ratios of gas to total mass in columns 4 and 5. Column 6 lists the infall rates (in $M_\odot yr^{-1}$) for the whole disk predicted at the present epoch: models with shorter infall e-folding times have current gas accretion lower than models with longer $\theta$. The last column gives the predicted helium variation with metallicity, which in most cases turns out to be consistent with the $\Delta Y/\Delta Z \simeq 4$ currently derived from observations.

Table 2. Output Parameters of the *Best* Models.

| Author | SFR($R_\odot, t_\odot$) | SFR($R_\odot$,T) | g/m($R_\odot, t_\odot$) | g/m($R_\odot$,T) | infall(T) | $\Delta Y/\Delta Z$ |
|---|---|---|---|---|---|---|
| Carigi | 5.8 | 2.8 | 0.27 | 0.15 | 0.24 | 2.5 |
| Ferrini | 2.2 | 1.0 | 0.28 | 0.18 | 0.15 | 2.4 |
| Matteucci | 2.9 | 1.3 | 0.16 | 0.08 | 1.00 | 1.6 |
| Prantzos-a | 6.8 | 4.0 | 0.31 | 0.18 | 0.28 | - |
| Timmes | ? | 4.1 | ? | 0.10 | 0.23 | 4 |
| Tosi-1 | 10.1 | 7.5 | 0.18 | 0.06 | 1.81 | 3 |

In sections 2 and 3, the most relevant differences resulting from the comparison of all the above *best* models are presented and discussed; common results with particular relevance for stellar and Big Bang nucleosynthesis are described in the last section.

## 2. Gas density and star formation

Fig.1 shows the local evolution of the SFR and the gas fraction predicted by the models listed in Tables 1 and 2. Tosi's models begin with the disk containing a finite amount of gas (see Fig.6), whereas the others assume that the disk building starts at t=0. In all cases it is assumed that the disk initially contained no stars, therefore $\Sigma_g/\Sigma_{tot}=1$ at t=0. The gas fractions at subsequent epochs depend on the different balancing between gas accretion and astration, but it is striking to see how similar to each other they result from these models. The predictions for the present local gas fraction all end up in the range 0.05−0.20, in excellent agreement with the data.

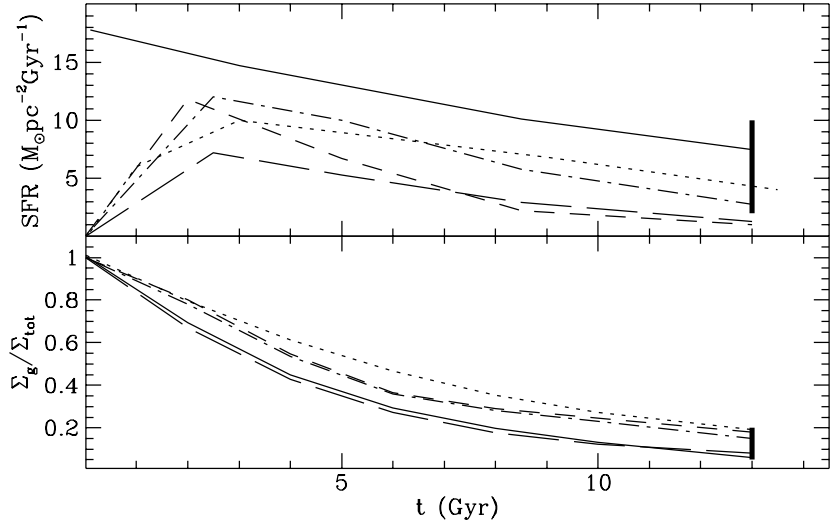

Figure 1. SFR (top panel) and gas fraction (bottom panel) predicted in the solar neighbourhood by Table 1 models. Symbols in both panels correspond to: dash-dotted line, Carigi; short-dashed line, Ferrini; long-dashed line, Matteucci; dotted line, Prantzos-a; solid line, Tosi-1. The vertical bars at t=13 Gyr show the range of values derived from local observations, as given by Timmes et al (1995).

Also the predicted star formation histories (top panel) look rather similar to each other, with the exception of Tosi's which is monotonically decreasing with time as a consequence of the adopted law (see Table 1). In all the other models the gas density reaches a maximum after a while (depending on the balancing between infall and SFR time scales) and then declines when the effect of gas consumption overcomes that of accretion. Since they assume a SFR related to the gas density, the time behaviour of their SFR follows that of $\Sigma_g$. The interesting point is that both the epoch of the peak and the following decline of the SFR do not differ much from one model to another and that all the predictions for the present SFR are roughly consistent with the observed range.

Fig.1 suggests that the results for the solar neighbourhood evolution obtained by the examined authors are consistent with each other, despite the different assumptions and approaches. The same conclusion can be reached comparing the predicted age-metallicity relation and G-dwarf frequency distribution with metallicity, which are also consistent with the corresponding observed data (see references to Table 1). However, we will see below that significant differences turn out to be present in the predictions of these models for the radial distributions of various quantities. This shows how important is to model different disk regions and to avoid to restrict to the solar neighbourhood the comparison between theoretical predictions and observational constraints.

For instance, the various predictions on the current distribution with galactocentric radius of both the SFR and the gas density (see respectively, top and bottom panel of Fig.2) show interesting differences. All the computed gas densities fall within the observed range of values (bottom panel), but this is due to the extreme uncertainty affecting the data. Three models (Matteucci's, Prantzos' and Tosi's) predict a gas density peak in the inner disk regions followed by a monotonic decrease with galactocentric distance, whereas in Carigi's and Ferrini's models the peak occurs well beyond the solar circle.

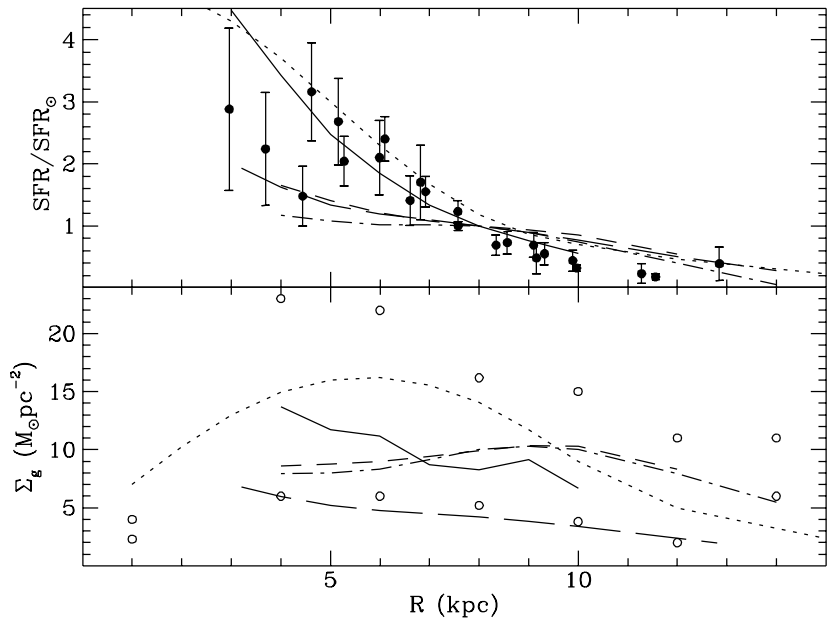

Figure 2. Radial distribution of the SFR (normalized to the value at the solar radius, top panel) and gas density (bottom panel) predicted for the present epoch by Table 1 models. Line symbols are as in Fig.1. Dots and error bars in the top panel correspond to observational data on recent SFR as presented by Lacey & Fall (1985); the open circles in the bottom panel delimit the range of observed gas densities adopted by the various modellers.

The recent star formation activity in various disk regions, as inferred from observations of young objects like O-stars, pulsars, Ly-continuum photons, is plotted in the top panel of Fig.2 and shows a steep decline with galactocentric distance. The predicted radial distribution of the star formation is intimately related to the gas density distribution, specially for models assuming the SFR proportional to $\Sigma_g$. In fact, the models (long-dashed, short-dashed and dash-dotted lines) with flatter $\Sigma_g$ profiles also show flatter SFR profiles. The only curves with SFR gradient steep enough to be in agreement with the observed trend are the dotted and solid lines. This is because Prantzos and Tosi adopt SFRs strongly declining outwards (see Table 1). In his case, the SFR is simply inversely proportional to R; in mine, it is proportional through the coefficient A to the total mass density, which does steeply decrease with galactocentric distance. Carigi and Matteucci also assume an explicit dependence of the SFR on $\Sigma_{tot}$, but with an exponent (0.4 and 0.1, respectively) too small to compensate the effect of the stronger dependence on $\Sigma_g$, which has instead a flat radial distribution. This figure shows that, whatever the actual mechanism triggering

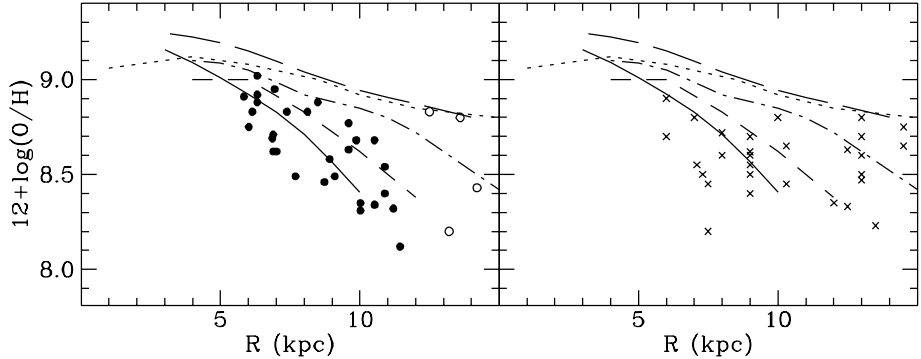

Figure 3. Radial distribution of the oxygen abundance at the present epoch as predicted by Table 1 models and as derived from observations of young objects (H II regions in the left panel: dots from Shaver et al 1983, circles from Fich & Silkey 1991; B stars from Prantzos & Aubert 1995 in the right panel). Line symbols are as in Fig.1.

the stellar formation, it must be more effective in the inner than in the outer galactic regions.

## 3. Abundance gradients

One of the most useful observational constraints for chemical evolution models is the radial distributions of the element abundances. From the published model results and/or the material provided directly by the authors it has been possible to compare the various predictions obtained for D, $^3$He, $^{12}$C, $^{13}$C, $^{14}$N, $^{16}$O, $^{17}$O, $^{18}$O and $^{56}$Fe.

Fig.3 shows the oxygen radial distribution predicted by the models at the present epoch together with the corresponding observational data. The overall theoretical distributions do not differ much from each other but some models show steeper gradients than others. The traditional empirical counterpart for the current oxygen abundances are the values derived from H II region observations (left panel) which show a rather strong radial decline ($\Delta \log(O/H)/\Delta R \simeq -0.09$ dex/kpc) up to approximately 12 kpc from the center and perhaps a flattening in the outermost regions. The models with steeper oxygen gradients are in better agreement with the distribution derived from H II regions: the model predictions represented by the dotted and the long-dashed lines appear too flat and overabundant.

However, it has been recently suggested that the errors in the abundance determinations in H II regions are much larger than usually quoted ($\pm 0.2$ dex) and that B stars are safer indicators of the actual oxygen distribution in the disk (right panel in Fig.3). Several people working on abundance determinations (e.g. Danziger 1995, Vilchez 1995, private communications) are skeptical about this possibility, but if it turns out to be true, then the oxygen gradient would be very shallow ($\Delta \log(O/H)/\Delta R \simeq -0.03$ dex/kpc) and only the two flatter model curves would have the appropriate slope (though still overabundant).

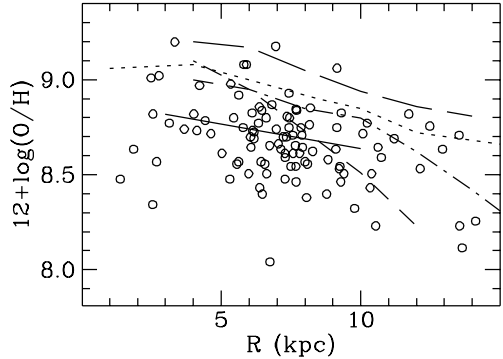

Figure 4. Radial distribution of the oxygen abundance 3 Gyr ago as predicted by Table 1 models and as derived from observations of PNeII (Pasquali & Perinotto 1993). Line symbols are as in Fig.1.

Fig.4 shows the oxygen radial distribution in the disk 3 Gyr ago. The data points are the abundances derived by Pasquali & Perinotto (1993) from a large sample of PNe of type II, which are supposed to have stellar progenitors born around that epoch. The abundance distribution independently derived by Maciel & Koppen (1994) for PNe II is in agreement with that shown in the figure. The flatter slopes are in better agreement with the data. Notice that Tosi's gradient is predicted to be flatter 3 Gyr ago than now (compare Fig.3 and Fig.4), Matteucci's and Prantzos' look roughly the same at the two epochs, Carigi's and Ferrini's are flatter now. This suggests that different models provide different histories of the abundance gradients.

To verify this result in more detail, in Fig.5 I have plotted as a function of time the slopes of all the element gradients derived by each author. In the last several billion years all the elements show negative gradients, with the exception of deuterium (dash-dotted lines) which is immediately burnt in the interior of any star and therefore is always more efficiently depleted in inner than in outer disk regions, because the SFR is and has always been higher toward the center. The absolute values and the trend with time of the gradient slopes vary from panel to panel. The reason for such different time behaviours of the metallicity gradients is that they depend on the model evolutionary scenario. Indeed the distribution with galactocentric distance of any element depends slightly on the time scale of its stellar production/ejection and mostly on the radial variation of the balancing between ISM enrichment from stars and ISM dilution from metal poor gas. In other words, the gradients result from the radial variation of the SFR/infall ratio. Regions with higher SFR always have larger chemical enrichment, but the polluting efficiency of such enrichment depends on the metal content of the ISM. If a region contains or accretes a large amount of metal poor gas its metallicity can remain relatively low even if the star formation is high.

To better understand the effect of the disk evolutionary scenario on the history of the abundance gradients it is useful to compare the different conditions resulting from different models. Fig.6 shows the radial distributions at various epochs of the SFR and the gas density in the two models for which I have outputs available for larger time intervals: Tosi's (left panels) and Prantzos' (right panels). My disk models start (dotted lines) with a finite, radially decreasing gas density. The initial SFR is radially decreasing as well, as a consequence of its dependence on the present gas and total mass densities. In these conditions, there is so much more metal free gas in the inner than in the outer regions that the ISM enrichment due to the first stellar ejecta is more efficient outside than inside, despite the larger SFR of the inner regions. As a consequence the gradients develop with positive slope. In Prantzos' case, at t=0 $\Sigma_g$=SFR=0. Soon after, gas accretion builds the disk with a radially decreasing gas density triggering a much more steeply decreasing SFR (see e.g. the dotted line for the situation at t=1.5 Gyr). With so much larger SFR in the inner regions the gradients develop immediately with negative slope.

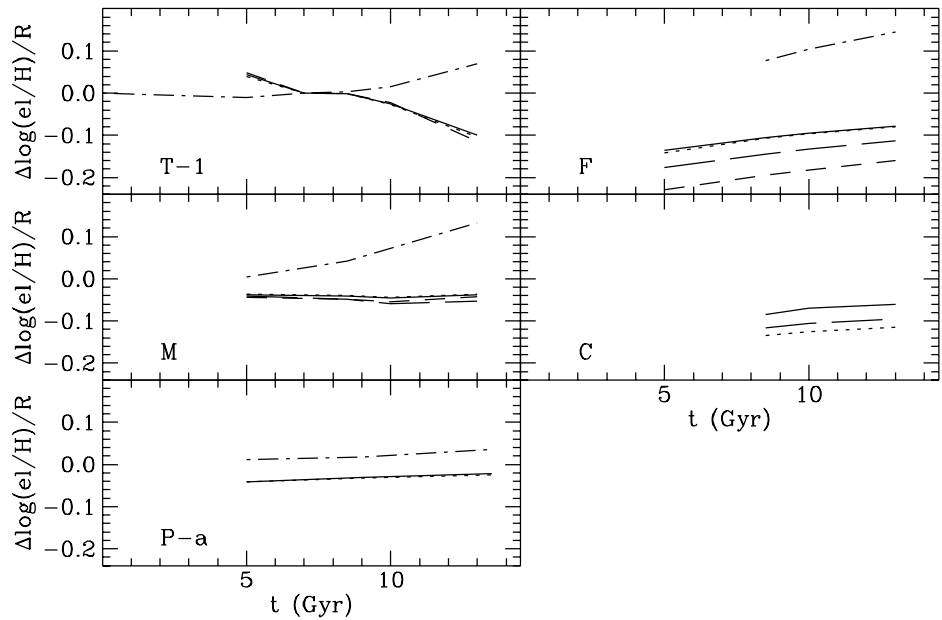

Figure 5. Time behaviour of the abundance gradients according to different authors (clockwise from the bottom left panel: Prantzos, Matteucci, Tosi, Ferrini and Carigi). Symbols are: dash-dotted lines for D; dotted for C; short-dashed for N; solid for O; long-dashed for Fe.

As time goes by, the disk conditions in the two models change significantly as a result of the different assumptions on the SFR and infall rate (recall that my models assume constant infall, his an infall e-folding time of 3 Gyr). At the epoch of the sun formation (dashed lines), Tosi's gas density is everywhere lower than at the beginning and still radially decreasing outwards, and the SFR is also lower. However, the effect of the long term, equidense infall is to bring much more metal poor gas to the outer than to the inner regions thus favoring the development of negative gradients. These conditions persist and reinforce in the subsequent epochs thanks to the higher efficiency of the infall dilution in the outer regions, and the gradients become more and more negative. In Prantzos case, instead, at t=8.5 Gyr the combined effect of the previous high star formation activity and the short time scale of the accretion rate has significantly depleted the amount of gas available in the inner regions, making the radial distribution of $\Sigma_g$, and consequently of the SFR, rather flat. In these conditions the efficiency of the ISM enrichment is low. The further flattening of the SFR radial profile at subsequent epochs and the almost complete gas consumption in the most central regions lead these ones to metallicity saturation, and the gradients flatten.

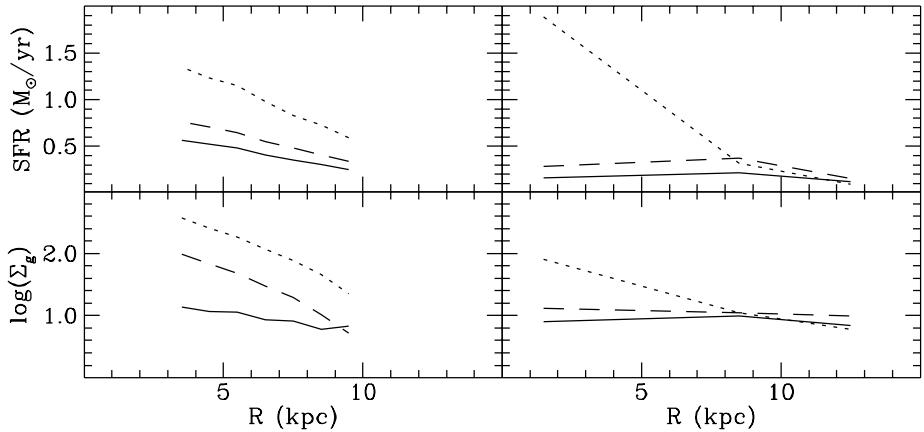

Figure 6. Radial distribution of the SFR (top panels) and the gas density (bottom panels) predicted at various epochs with the *best* models by Prantzos (right panels) and Tosi (left panels). Symbols are: dotted lines for the earliest distributions, (t=0 for Tosi-1 and t=1.5 Gyr for Prantzos-a); dashed for t=8.5 Gyr; solid for present epoch.

It is then clear that the history of the metal abundance gradients results from the SFR and infall rate behaviours with time and space. In turn, if reliable data were available for the metallicity gradients both at recent and at early epochs, we could use them as good indicators of the star formation and gas accretion histories in the disk. Unfortunately, despite their accuracy, the data obtained from open clusters and field stars observations are not sufficient yet to discriminate between the various scenarios. For instance, the (intrinsic) dispersion in the metallicity distribution with age and galactocentric distance of the Edvardsson et al (1993) sample is so large that the corresponding curves predicted by Carigi, Matteucci, Prantzos and Tosi, all fall well within the observational range, despite the different scenarios of these models.

## 4. Interesting elements and element ratios

The model predictions on D, $^3$He and C and O isotopes deserve further discussion, for their implications on cosmology and stellar nucleosynthesis.

The two lighter elements provide fundamental clues on the theory of Big Bang nucleosynthesis, since they are produced directly there (together with H, $^4$He and $^7$Li) in amounts which depend on the, yet unknown, baryon to photon ratio $\eta$. The left panels in Fig.7 show the local evolution of the D and $^3$He abundances by mass predicted by Table 1 models from the same primordial values. The agreement in the D behaviour (top panel) is striking: all the models fit the observed ranges and all of them predict a D depletion factor <3 from the primordial to the present value. Since these models (and only them) are consistent with the whole set of observational constraints, in and outside the solar neighbourhood, this common result is a fairly stringent limit on the D depletion factor. Given the observed D values in the solar system and in the local ISM, this implies a primordial $X_2 < 8 \times 10^{-5}$. More exotic models have been invoked by other authors to allow for larger D consumptions, but **none** has been compared with all the available constraints. If such comparisons were performed those models would probably result in conflict with many constraints.

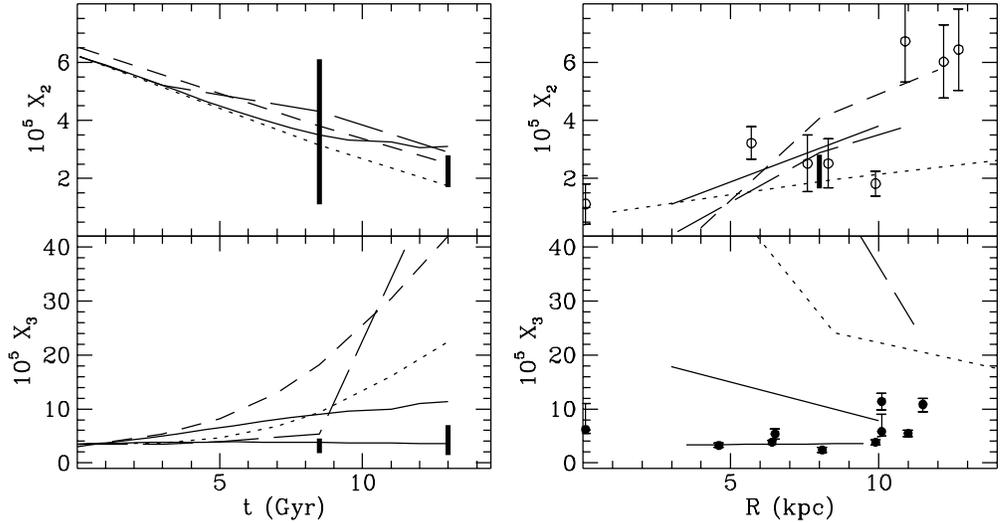

Figure 7. Left panels: Time behaviour of the D (top) and $^3$He (bottom) mass fractions predicted in the solar neighbourhood by Table 1 models; the vertical bars show the 2-$\sigma$ ranges of values derived from local ISM and solar system observations (Dearborn et al 1996). Right panels: Corresponding radial distribution at the present epoch; the vertical bar at 8 kpc is as in the top left panel, open circles are Wannier's (1980) data divided by 100, filled circles Rood et al's (1995) data. Line symbols are as in Fig.1; the lower solid line in the bottom panels shows the effect of Hogan's deep mixing on the $^3$He yield.

For $^3$He (bottom panels in Fig.7), instead, all the models are inconsistent with the observed abundances, but the interesting point is that this is due to one common cause: the excessively large $^3$He production in low and intermediate mass stars which leads to an ISM overenrichment. This large production is predicted by all the standard nucleosynthesis models (see references in Galli et al 1995, Dearborn et al 1996 and Prantzos 1996) but there are several attempts to circumvent it through the large $^3$He depletion which would occur in stellar envelopes in case of deeper mixing of the convective zones (e.g. Hogan 1995, Charbonnel 1996). These *non standard* models are difficult to reconcile with the large $^3$He content observed in a few PNe with stellar progenitors of presumed initial mass around 1.5 M$_\odot$ (see Dearborn et al 1996); however if they eventually turn out to be valid, the effect on the evolution of $^3$He would be to significantly reduce its ISM abundance, as shown for instance by the lower solid line in the left bottom panel, which fits well the observational data.

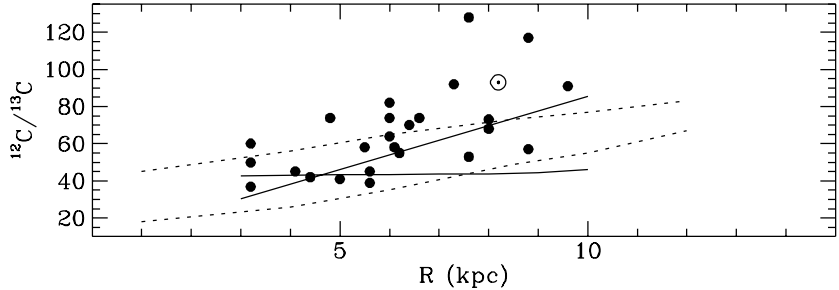

Figure 8. Radial distribution of $^{12}C/^{13}C$ at the present epoch as derived from molecular cloud observations (Henkel et al 1985) and as predicted by Prantzos (dotted lined) and Tosi (solid) models.

The same conclusions on both D and $^3$He can be reached from the current radial distributions of the two elements, plotted in the right panels of Fig.7. For D, the abundances derived from Wannier's (1980) observations of molecular clouds are added to the local ISM data to show what may be its observed radial trend. To take into account the possible effect of chemical fractionation in deuterated molecules, Wannier's data have been divided by 100.

The evolution of the C and O stable isotopes predicted by the various *best* models also indicates that something should be revised in stellar nucleosynthesis. In fact, the distributions of $^{12}C/^{13}C$ and of $^{16}O/^{18}O$ with time and galactic radius predicted on the basis of standard nucleosynthetic assumptions are inconsistent with the corresponding data. For $^{12}C/^{13}C$ (Fig.8) the problem is that the present ratios predicted by models based on Renzini & Voli's (1981) yields are lower and/or less steeply distributed with galactic radius than the observed ones. A steeper gradient is reached by assuming for $^{13}C$ either a delayed ejection consequence of a major production in Nova events (e.g. D'Antona & Matteucci 1991) or a completely secondary nature (e.g. Tosi 1982), but this does not always imply also consistent absolute values for the present ratio (e.g. Prantzos et al 1996).

For the oxygen isotopes the problem is that the predicted $^{16}O/^{18}O$ and $^{17}O/^{18}O$ ratios in the ISM are respectively higher and lower by a factor 1.6 than those inferred from molecular cloud observations. The issue is puzzling; however, since both $^{16}O$ and $^{16}O/^{17}O$ show time and space behaviours in agreement with the model predictions, it is most probable that what causes the disagreement is $^{18}O$. Years ago it was suggested (Tosi 1982) that the problem would be solved if the amount of $^{18}O$ ejected by stars can decrease after the epoch of the solar formation. This possibility has not yet been examined with stellar nucleosynthesis models; however the yields metallicity dependence recently demonstrated by Maeder (1992) may provide a good motivation for it. Alternative ways out have been suggested by Prantzos et al (1996): either the observational $^{16}O/^{18}O$ is overestimated by a factor 1.6, or the solar system ratio is not representative of that of the local ISM 4.5 Gyr ago.

To conclude, the models currently in better agreement with the majority of the observational constraints show a general agreement in the predictions for the solar neighbourhood evolution and in the implications for stellar nucleosynthesis, but predict fairly different scenarios for the history of different disk regions because of their different assumptions on infall and SFR.

**Acknowledgments.** I wish to thank L.Carigi, C.Chiappini, F.Matteucci and N.Prantzos for having made this comparison possible by running their models for me and by sending the details of their most recent results. M.Mollà has also helped providing unpublished information on Ferrini et al's models.